\documentstyle[prb,aps,eqsecnum,epsf]{revtex}
\begin{document}
\title{\LARGE Multiquantum Vortices in Conventional Superconductors with 
Columnar Defects Near the Upper Critical Field}
\author{Gregory M. Braverman, Sergey A. Gredeskul, 
and Yshai Avishai}
\address{Department of Physics, Ben Gurion University of the Negev, Beer Sheva 
84105 ISRAEL}
\date{February 24, 1998}
\maketitle
\begin{abstract}
Equilibrium vortex configuration in conventional type II superconductors 
containing  columnar defects is theoretically investigated.
Near the upper critical field a single defect causes a strong local 
deformation  
of the vortex lattice. 
This deformation has $C_3$ or $C_6$ point symmetry, whose character 
strongly depends on the vortex-defect interaction. If the interaction is 
attractive, the 
vortices can collapse onto defect, while in the case of 
repulsion the regions free of vortices appear near a defect. Increasing  
the applied magnetic field results in an abrupt change of the configuration 
of  
vortices giving rise to reentering transitions between configurations 
with $C_3$ or $C_6$ symmetry. In the case of a small concentration of  
defects these transitions manifest themselves as jumps of magnetization 
and discontinuties of the  magnetic susceptibility.
\end{abstract}
\begin{center}
\end{center}
\large
\section{Introduction}
Mixed state or Shubnikov phase \cite{Shubnikov} of type II 
superconductors is characterized by penetration of vortices into 
the sample \cite{Abrik} each one carrying the superconducting 
flux quantum $\phi_S=\pi \hbar c/e$. A single vortex has a normal core with 
radius 
of order $\xi (T)$ (the coherence length at temperature $T$)
surrounded 
by a closed superconducting current occupying the tube with 
radius $\lambda(T)$ 
(the penetration length). Near the upper
critical field $H_{c2}$ these vortices form a triangular Abrikosov
lattice \cite{Abrik}. If an external current is applied, vortices
start to move due to the Lorentz force. This motion leads to an energy
dissipation.
Different kinds of defects, such as dislocations, point defects or 
regions with different superconducting properties create some additional 
field acting on the vortices. As a result
vortices are pinned and nondissipative current of finite amplitude can flow 
through the superconducting sample. There are two different types of 
pinning \cite{LOreview}. Weak
defects lead to the so-called {\it collective pinning} \cite{LO1}. 
In this regime the vortex lattice is slightly deformed. This deformation
is well described by elasticity theory \cite{Labusch,L,Brandt1,LO1}.
Strong defects lead to a {\it single-particle pinning} \cite{LO1}.
A single strong defect is able to pin a vortex and, at a finite defect 
concentration, formation of metastable states is possible \cite{Ovch}.
A detailed theory of 
vortex pinning in conventional type II superconductors was formulated by 
Larkin and Ovchinnikov (see review paper \cite{LOreview}).\\

The discovery of high-temperature superconductivity \cite{BedMul} resulted in 
a deep 
understanding of the new rich and fundamental properties of vortex systems 
(see 
an exhaustive review papers of Blatter {\em et. al.} \cite{Blatter} and 
Brandt \cite{BrandtRev} and references therein). 
Statistical mechanics of vortices was formulated and new concepts appeared 
such 
as melting of the vortex lattice, vortex liquid and vortex glass. 
 The usage of heavy ion irradiation for preparation of 
superconducting samples with columnar defects  \cite{civale} 
opened new experimental possibilities to study the properties of 
vortex matter. 
Columnar defects serve as strong pinning centers, each of which is 
able to pin a single vortex as a whole. The radius of the columnar defect 
could be less than the Abrikosov 
lattice constant $a$ near $H_{c2}$. Such defects are referred to as 
short-range ones.\\

Strong columnar defects with radius $L$ much larger than 
the coherence length may lead to the formation of multiquantum vortices 
 in high temperature superconductors \cite{Buzdin}. Such vortices were 
observed experimentally  on submicron artificial holes in 
mutlilayers $Pb/Ge$ \cite{Moshchalkov}. Multiquantum vortices can  also be formed at large pinning
 centers with radius of order of the penetration length \cite{Shapiro}.   
In this paper we show that columnar defects can also strongly affect the  
properties of conventional type II superconductors.  In such 
superconductors near the upper critical field even the 
short-range columnar defects cause a strong local deformation of the vortex 
lattice due to its softening and as a result to the formation of multiquantum 
vortices.\\  

In the main part of the paper we consider a 
superconductor containing  a  single short-range columnar defect. 
When the magnetic field approaches to $H_{c2}$ the strength of the defect 
effectively increases resulting in 
strong lattice deformation in its vicinity. Initially the  Abrikosov lattice is 
triangular. Therefore  
the local deformation belongs to one of the two possible symmetry 
types -- $C_6$ or $C_3.$ In the case of an attractive defect,  
the vortices can collapse onto this defect with increasing of a magnetic field. 
As a result, reentering transitions between two local 
symmetries are possible. For example, at some external field the 
local vortex configuration $C_6$ with a single vortex pinned by a defect is 
preferred over the $C_3$ configuration with the defect placed at the center 
of a  triangle. But with increasing of a magnetic field the  
closest three vortices in the $C_3$ configuration could collapse onto the 
defect and a $C_6$--$C_3$ 
transition occurs. Now the mostly preferred local configuration is of 
$C_3$ type, with a three-quantum vortex on the defect. Further increasing of a 
magnetic field results in a $C_6$ type configuration with a seven-quantum 
vortex at the defect and so on. 
In the case of a small concentration of defects 
(which was realized in an experiment \cite{civale}, where 
the radius of the defect is equal to $2.5$ {\em nm} and the average 
distance between
defects is $4,600$ {\em nm}).
These transitions manifest themselves as jumps of magnetization and 
discontinuties of the magnetic 
susceptibility curve.\\      

The present paper is organized as follows. In the second section we 
formulate the problem. Further, in the third section we study the
relatively simple case of small deformation of the 
vortex lattice. It is realized for weak defects or for values of the 
magnetic field which are not very close to $H_{c2}$. The results of this 
study enable 
us 1) to scale the defect parameters with the magnetic field, and 2) to 
predict 
the occurrence of symmetry change when the applied magnetic field increases. 
The central section IV containes the results of the numerical solution of 
the pertinent equations. Here we present a universal phase diagram of the 
superconductor near the upper critical field, and analyze the vortex lattice 
deformation as a 
function of the magnetic field for various defect parameters. 
The next section V is devoted to the case of small concentration of the 
defects. Here we estimate the high order concentration corrections with respect 
to defect and study the magnetization and the magnetic susceptibility 
behavior near the upper critical field. 
Section VI summaries the main results. In the Appendix, the Abrikosov
lattice 
expansion in terms of the first Landau level wave functions is obtained for 
an arbitrary position of the lattice with respect to the origin.\\
\section{Formulation of the Problem}
Consider a  superconductor containing columnar defects and subject to an 
external magnetic field ${\bf H}=H\hat{z}.$ Both the defect column axis 
and the magnetic field are assumed to be directed along the $z$-axis. The 
unit volume thermodynamic potential of such a 
superconductor at a fixed temperature $T$ close to 
the critical temperature $T_c$ can be written as
\begin{equation}
F=\frac{1}{S}\int f\left({\bf r};
\left[\alpha,\gamma,\Psi,{\bf A}\right]\right) d{\bf r},        
	\label{neqfe}
\end{equation}
where ${\bf r}$ is the two dimensional (2D) position vector and the 
Ginzburg--Landau 
density 
$f$ of the thermodynamic potential \cite{GL} is
\begin{equation}
f\left({\bf r};
\left[\alpha,\gamma,\Psi,{\bf A}\right]\right)  
	=\alpha ({\bf r})|\Psi({\bf r})|^2+
	\frac{\beta}{2}|\Psi({\bf r})|^4+
 \gamma({\bf r})\left|{\bf\partial}_-\Psi({\bf r})\right|^2
 +\frac{1}{8\pi}({\bf B ({\bf r})}-{\bf H})^2.
	\label{dneqfe} 
\end{equation}
Here $\Psi ({\bf r})$ and ${\bf A}({\bf r})$    
are the order parameter and 2D vector potential respectively, 
$${\bf B}=\hat{z}\left(\frac{\partial A_y}{\partial x}-
\frac{\partial A_x}{\partial y}\right),$$ 
and $\partial_-$ is the gauge 
invariant gradient
\begin{equation}
{\bf\partial}_-\equiv -i\hbar \frac{\partial}{\partial {\bf r}}+
\frac{2e}{c}{\bf A}.
\label{gradin}
\end{equation}
The space dependent Ginzburg--Landau coefficients $\alpha ({\bf r})$ and 
$\gamma ({\bf r})$ for defects placed at the
points 
$\{ {\bf r}_j \}$ have the form
$$\alpha ({\bf r})=\alpha_0 +\sum_{j}\alpha_1 ({\bf r}-{\bf r}_j),$$
$$\gamma({\bf r})=\gamma_0 +\sum_{j}\gamma_1 ({\bf r}-{\bf r}_j),$$
where $\alpha_0<0$ and $\gamma_0>0$ correspond to a uniform superconductor 
and the short-range functions $\alpha_1({\bf r})$ and $\gamma_1({\bf r})$ 
describe the perturbation of these coefficients caused by a columnar defect 
located at the origin.\\

According to the standard procedure one has to minimize the 
density (\ref{dneqfe}) (i.e. to solve the Ginzburg--Landau equations), 
to find the extremal order parameter $\Psi$ and vector potential ${\bf A}$ 
and to substitute them into the formula (\ref{neqfe}). Assume now that the 
density $n$ of defects is small, $n\xi^{2}(T)\ll 1$, i.e. the average
distance 
between defects $n^{-1/2}$ is much 
larger than the coherence length $\xi(T)$ (which is of order of the 
distance $a$  
between neighboring vortices of the Abrikosov lattice). In this case 
the concentration expansion \cite{Lifshits} of the thermodynamic 
potential density in linear approximation yields
\begin{equation}
	F=f^{A}+n\int (f_1({\bf r})-f^{A}) d{\bf r}.
	\label{conc}
\end{equation}
Here 
\begin{equation}
	f^{A}=-\frac{1}{8\pi}\frac{(H-H_{c2})^2}{(2\kappa^2-1)\beta_A},
	\label{FE}
\end{equation}
is the free energy of the Abrikosov triangular lattice, 
$\beta_A=1.1596$ \cite{Kleiner} and 
$f_1({\bf r})$ is the minimum of the density (\ref{dneqfe}) 
containing a single defect placed at the origin 
$(\alpha({\bf r})=\alpha_0+\alpha_1({\bf r}),\phantom{aa}  
 \gamma({\bf r})=\gamma_0+\gamma_1({\bf r}))$.
Near the upper critical field 
$$H_{c2}=\frac{|\alpha_0|c}{2\gamma_0 \hbar e},$$
corresponding to a uniform superconductor with 
$\alpha ({\bf r})=\alpha_0,$ $\gamma({\bf r})=\gamma_0,$ the minimization 
procedure can be applied to the density of the thermodynamic 
potential
\begin{equation}
    f_1\left({\bf r};\left[\alpha,\gamma,\Psi\right]\right)=
    \alpha_0\left(1-h\right)|\Psi|^2
 +\frac{\beta}{2}\left(1-\frac{1}{2\kappa^2}\right)|\Psi|^4
 +\alpha_1({\bf r})|\Psi|^2 
 +\gamma_1({\bf r})|{\bf\partial}_{-}^{0}\Psi|^2,       
	\label{dneq}
\end{equation}
which depends only on the order parameter \cite{Brandt2,LO1}. Here 
$h\equiv H/H_{c2},$ $\kappa$ 
is the Ginzburg--Landau parameter 
$$
\kappa=\frac{c}{4\hbar e \gamma_0}\sqrt{\frac{\beta}{2\pi}},
$$ 
and $\partial^{0}_{-}$ is defined by Eq.(\ref{gradin}) where the 
vector potential ${\bf A}_0$ of an applied field ${\bf H}$ stands for
 ${\bf A}$. 
In what 
follows we will use the vector potential in the symmetric gauge 
${\bf A}_0= H (-y/2,x/2)$.\\

 To find 
 the order parameter which realizes this minimum one can use an expansion 
 of $\Psi$ 
 in terms of Landau functions $L_k(x,y)$ (\ref{Land}) of 
 the lowest Landau level of 
 a particle with electron mass and the charge $-2e$ in the magnetic 
 field $H,$
 substitute this expansion into Eq.(\ref{dneq}) and find the expansion 
 coefficients from the minimum condition \cite{Ovch}. Such an expansion serves 
 as a good approximation and one can neglect the contribution of the 
 highest Landau levels even at a field $H=0.5H_{c2}$ \cite{Brandt}.
 In linear 
 concentration approximation the problem is reduced to a single defect 
 problem. Therefore in the case of 
 isotropic functions $\alpha_1(r)$ and  $\gamma_1(r)$ the symmetry 
 of the unperturbed Abrikosov lattice enables us to consider only two 
 cases corresponding either to $C_6$ symmetry, or to $C_3$ symmetry. The 
 hexagonal symmetry corresponds to the distorted vortex lattice with 
 one vortex placed on the defect. The trigonal one corresponds to the 
 lattice with the defect located in the center of the vortex triangle.  
 In the hexagonal case the trial order parameter can be written as
 \begin{equation}
	\Psi_{6}(x,y)=i\sum_{k=0}^{\infty}
  \left[\pi^{-1}M_6(k)+D(k)\right]
  L_{k}(x,y).
	\label{C6}
 \end{equation}
 Here $D(k)$ are the variational parameters which should be found. The case 
 when all $D$ are equal to zero and only the  
 coefficients $M_6(k)$ remain, corresponds to the order parameter 
 $\Psi_{6}^{A}(x,y)$  which describes the Abrikosov lattice with 
 one of the vortices located at the origin and one of the symmetry axes 
 parallel to the $x$-axis. The coefficients $M_6(k)$ (see 
 Eq.(\ref{M6}) of Appendix) are real and   
 obey the selection rule \cite{Ovch} $k=6K+1,\phantom{aa} 
 K=0,1,2, ...\phantom{aa}.$ 
 In the trigonal 
 case the trial order parameter is written as
 \begin{equation}
	\Psi_{3}(x,y)
  =\sum_{k=0}^{\infty}i^{-k}\left[\pi^{-1}M_3(k)+ D(k)\right] L_{k}(x,y).
	\label{C3}
 \end{equation}
 The case when all $D$ are equal to zero, corresponds to the order parameter
  $\Psi_{3}^{A}(x,y)$  which describes the Abrikosov lattice whose origin 
 coincides with the center of the vortex triangle and  one of the symmetry axes 
is parallel to the $x$-axis. The real coefficients  $M_3(k)$ (\ref{M3}) obey 
 the selection rule $k=3K, K=0,1,2, ...\phantom{aa}.$\\
 
 Thus to obtain the lattice deformation caused by a single defect we have to 
 find separately the extremal set of the 
 variational parameters $D(k)$ within each of the two 
 symmetry classes separately, and to choose the most preferable one from the 
 two of them. 
 This procedure and its consequences will be discussed in the next two 
 sections.\\
\section{Weak Defects} 
Consider first a system with weak defects (in a sense that will be clear 
later on). It is 
natural to assume that in this case the two last terms in the thermodynamic 
potential density (\ref{dneq}) 
do not contribute to the variational equation for the order parameter and 
the latter one coincides with its Abrikosov value $\Psi^{A}.$ 
Accordingly, the equilibrium thermodynamic potential can be written as 
\begin{equation}
	F=f^{A}+n\int (\alpha_1({\bf r})|\Psi^{A}|^2 
 +\gamma_1({\bf r})|{\bf\partial}_{-}^{0}\Psi^{A}|^2) d{\bf r}. 
	\label{conc1}
\end{equation}
Let us then specify the functions $\alpha_1({\bf r})$ and 
$\gamma_1({\bf r})$ 
which describe the perturbation of the Ginzburg-Landau coefficients by defects
$$\alpha_1({\bf r})=-\alpha_0\tilde{\alpha}
 \exp\left(-\frac{r^2}{2L^2},\right)$$ 
 $$\gamma_1({\bf r})=\gamma_0\tilde{\gamma}
 \exp\left(-\frac{r^2}{2L^2}\right).$$
Here $\tilde{\alpha}$ and $\tilde{\gamma}$ describe the strengths of 
the defect, measured in units $\alpha_0$ and $\gamma_0$ respectively, 
and $L$ is its size. Accurate estimations show that if 
the properly scaled strengths of defects
\begin{eqnarray}
 \alpha=\frac{\tilde{\alpha}}{1-h}\nonumber\\
 \gamma=\frac{\tilde{\gamma}h}{1-h}
 \label{scale}
\end{eqnarray}
are small, $\alpha, \gamma \ll 1,$ then one can indeed neglect the Abrikosov 
lattice distortion.\\

It seems that in the attractive case the preferable configuration is 
always a $C_6,$ i.e. Abrikosov lattice with one of the vortices located on a defect. 
Nevertheless we will show that even in the case of small deformation, the 
previous statement is not always valid.  In the general case one should take 
into account the two possible types of lattice symmetry with respect 
to a given defect, i.e. the 
two Abrikosov order parameters $\Psi_{6,3}^{A}$ corresponding to the $C_6$ 
symmetry 
$$\Psi_{6}^{A}(x,y)=i\pi^{-1}\sum_{k=0}^{\infty} M_6(k) L_{k}(x,y),$$
and to the $C_3$ symmetry 
$$\Psi_{3}^{A}(x,y)
  =\pi^{-1}\sum_{k=0}^{\infty}i^{-k}M_3(k) L_{k}(x,y).$$\\
 
Substitution of these order parameters in the Eq.(\ref{conc1}) yields the 
corresponding thermodynamic potentials,
\begin{eqnarray}
 F_6=f^{A}+\frac{4\pi c}{3^{1/4}}\left|f^{A}\right|M_6^2(1)
 \gamma \varphi,\nonumber\\
 F_3=f^{A}+\frac{4\pi c}{3^{1/4}}\left|f^{A}\right|M_3^2(0)
 \alpha \varphi.
 \label{FE63}
\end{eqnarray}
Here $\varphi$ is the dimensionless scaled defect size
\begin{equation}
	\varphi=h\frac{L^2}{\xi^2(T)},
	\label{radius}
\end{equation}
$c=na^2\sqrt{3}/2$ is the dimensionless defect concentration (the number 
of defects per a single vortex) 
and $a$ is the Abrikosov lattice constant (see Eq.(\ref{latconst}) below).\\

Suppose we deal with a defect in which the only variation parameter is  
$\tilde{\alpha}$ (i.e. the transition temperature).
Then if $\tilde{\alpha}>0$ such  defect increases the thermodynamic potential
 $F_3$ leaving  the 
 $F_6$ unchanged (\ref{FE63}). This means that the defect attracts a 
 vortex and the symmetry $C_6$ is 
preferable. Obviously,
in the opposite case $\tilde{\alpha}<0$ the symmetry $C_3$ is preferable and
therefore the defect is repulsive. This confirms the qualitative speculations 
presented in the Introduction. But the question is what happens if variation 
of both $\tilde{\alpha}$ and $\tilde{\gamma}$ is allowed. To answer 
this question 
we must compare the correction terms in Eqs.(\ref{FE63}). 
Taking values of $M_6(1)$ and $M_3(0)$ from the Tables I, II of Appendix we 
conclude that if 
\begin{equation}
 \gamma\leq 0.387\alpha,
 \label{ineq}
\end{equation}
then the symmetry $C_6$ is preferable (attractive case). Yet, the ratio 
$\alpha/\gamma$ grows with the magnetic field (see Eq.(\ref{scale}) above).
Therefore if, for some field, $\gamma$ is slightly less than $0.387\alpha$ 
then further increasing of a 
magnetic field can violate the inequality (\ref{ineq}) and causes a first 
order phase transition to the $C_3$ symmetry.\\ 

In the region of field considered above, the vortex lattice is comparatively
 rigid 
and vortex repulsion dominates above vortex-defect interaction. However even in 
this region the type of the lattice symmetry can be changed. For stronger 
magnetic fields or for stronger defects the lattice  deformation near defect 
is not negligible any more. This leads to richer and more complicated 
properties of the vortex system even in the case when 
$\tilde{\gamma}=0$.\\

\section{Strong Deformation of the Vortex Lattice}
Now consider the case when a lattice deformation near defects is essential.
This deformation is completely  described by an infinite set of variational 
parameters 
$\{ D(k)\}$. Direct substitution of the test function $\Psi(x,y)$ 
 expressed in the forms (\ref{C6}) or  (\ref{C3})  into the 
expression for the thermodynamic potential  
(\ref{conc}), (\ref{dneq}) yields
\begin{equation}
 F=f^{A}\left[1-\frac{4\pi c}{3^{1/4}}Q\right],
\label{F}
\end{equation}
where
\begin{eqnarray}
 &&Q=\frac{2}{3^{1/4}\beta_A}\left\{
 \sum_{k,l,m}\frac{(l+m)!}{2^{l+m+2}\sqrt{k!l!m!(l+m-k)!}}\left[
 \pi D^{*}(k)D(l)D(m)D^{*}(l+m-k)+\right.\right.\nonumber\\
 &&\left.2M(l+m-k)(D(k)D^{*}(l)D(m)+c.c.)\right]+2\sum_{k,l}I(k,l)D^{*}(k)D(l)+
 \nonumber\\
 &&\left .\frac{1}{2}\sum_{k,l}\sqrt{\frac{(k+l)!}{k!l!}}J(k+l)(D(k)D(l)+c.c)
 \right \}+\sum_k |D(k)|^{2}+\nonumber\\
&&\sum_k
 |\pi^{-1}M(k)+ D(k)|^2
 \frac{\varphi ^k}{(1+\varphi)^k}
 \left[
 \alpha \varphi +\gamma  (\varphi^2+
 k(1+2\varphi ^2))\right],
\label{Q}
\end{eqnarray} 
and
$$I(k,l)=\sum_{m}\frac{(l+m)!}{\sqrt{k!l!m!(l+m-k)!}}
\frac{M(m) M(l+m-k)}{2^{l+m+1}\pi},$$
$$J(k)=\sum_{m}\sqrt{\frac{k!}{m!(k-m)!}}\frac{M(m)M(k-m)}{2^{k+1}\pi}.$$
This expression for the correction to the thermodynamic potential is general 
and valid for 
both two symmetries $C_6$ and $C_3$. In each of these cases one should 
take into account the selection rules
\begin{eqnarray}
&M_{3}(k)=\delta_{k,3K}M_{3}(3K),&\nonumber\\
 &M_{6}(k)=\delta_{k,6K+1}M_{6}(6K+1),&\nonumber\\
 &K=0,1,2, ...\phantom{a} ,&
\label{selectionM}
\end{eqnarray}
and  use for $M_{3,6}(k)$ their corresponding (real) values 
(see Eqs.(\ref{M6}), (\ref{M3}) below).
The next step is the minimization of the thermodynamic potential 
(\ref{F}), (\ref{Q}) with respect
to the coefficients $\{D(k)\}$. The equations which determine  $\{D(k)\}$ 
have the form
\begin{eqnarray}
 &&\frac{2}{3^{1/4}\beta_A}
 \left\{
  \sum_{l,m}\frac{(l+m)!}{2^{l+m+1}\sqrt{k!l!m!(l+m-k)!}}
  (\pi D(l)D(m)D^{*}(l+m-k)+M(l+m-k)D(l)D(m))\right.\nonumber\\
  &&\left.\sum_{l,m}
  \frac{(k+m)!M(k-l+m)}{2^{k+m}\sqrt{k!l!m!(k-l+m)!}}D(l)D^{*}(m)
+\sum_{l}
  \left[
   2I(k,l)D(l)+\sqrt{\frac{(k+l)!}{k!l!}}J(k+l)D^{*}(l)\right]\right\}\nonumber\\
 &&-D(k) +
 \left(\pi^{-1}M(k)+ D(k)\right)
 \frac{\varphi^{k}}{(1+\varphi)^k}
 \left[
 \alpha \varphi+\gamma(\varphi^2+k(1+2\varphi^2))\right]=0
\label{Ovch}
\end{eqnarray}
and were obtained by Ovchinnikov \cite{Ovch} who used their linearized 
version for studying possible structural transitions.\\
  
We numerically solve the  infinite nonlinear system of Ovchinnikov equations  
without any simplification. The only (quite natural and verified) assumption 
which we 
use is that the perturbed lattice conserves its initial symmetry. This 
means that the coefficients $\{D(k)\}$ obey the same selection 
rules 
\begin{eqnarray*}
&D_{6}(k)=\delta_{k,6K+1}D_{6}(6K+1),&\nonumber\\
&D_{3}(k)=\delta_{k,3K}D_{3}(3K),&\nonumber\\
&K=0,1,2, ... \phantom{a}.&
\end{eqnarray*}
that the initial coefficients $M(k)$ do. Our strategy is as follows. 
For fixed values of the parameters $\tilde{\alpha}$ and 
 $\tilde{\gamma}$ and for a fixed magnetic field we calculate the 
 coefficients $D(k)$ for two possible symmetries $C_6$ and $C_3.$ 
 Then we substitute these solutions together with the corresponding sets 
 of $\{M\}$ into Eq.(\ref{Q}) and choose the most preferable solution which 
 determine 
the vortex lattice deformation as well as the thermodynamics of the system 
to first order in the low concentration approximation. Thus to understand the 
results obtained we should first analyze the behavior of the coefficients 
$D(k)$ in a magnetic field and to explain how this behavior influences to 
the order parameter evolution within each of the two symmetries separately. 
Then we can describe the vortex configuration, corresponding to the preferable 
solution for a fixed set of parameters 
$\tilde{\alpha},\phantom{a}\tilde{\gamma},\phantom{a}L,$ 
and its evolution in a magnetic field.\\

The qualitative information concerning the behavior of the coefficients 
$D(k)$ in a magnetic field can be obtained directly from Eqs.(\ref{Ovch}). 
Consider for example an attractive defect with $\tilde{\alpha}>0$ and 
$\tilde{\gamma}=0.$ In this case, if one is not too close to the critical 
field $H_{c2}$ the hexagonal symmetry should be realized and one starts from 
an analysis of the $C_6$ solutions. Due to selection rules, the first 
nonvanishing equation of the system 
(\ref{Ovch}) will correspond to the value $k=1$.
This equation strongly depends on the (scaled) defect parameters 
$\alpha,$ $\gamma$ and $\varphi$, which are collected 
in the last term of Eq.(\ref{Ovch}). But right in the next equation (which 
corresponds to the value $k=7$) this term is proportional to $\varphi^{7}$ 
and due to the short range nature of the defect $(\varphi\leq 1$)  is very small. 
Therefore all the higher order equations (\ref{Ovch}) with $k=13, 19, 
...$ are practically homogeneous. As a result, the solution of 
(\ref{Ovch}) will give nonzero coefficients $D(k)$ only for some small 
values of $k$. 
Thus the deformation of a vortex lattice happens mainly near the defect, at 
the distance of order of the Larmor radius $R_k\propto \sqrt{k_{max}}$ \
corresponding to the largest value of $k$ such that $D(k_{max})\neq 0$, while 
the rest of the lattice remains undistorted.\\
\begin{figure}
  \begin{center}
  \epsfxsize=10.7cm
  \epsfysize=10.85cm
  \epsfbox{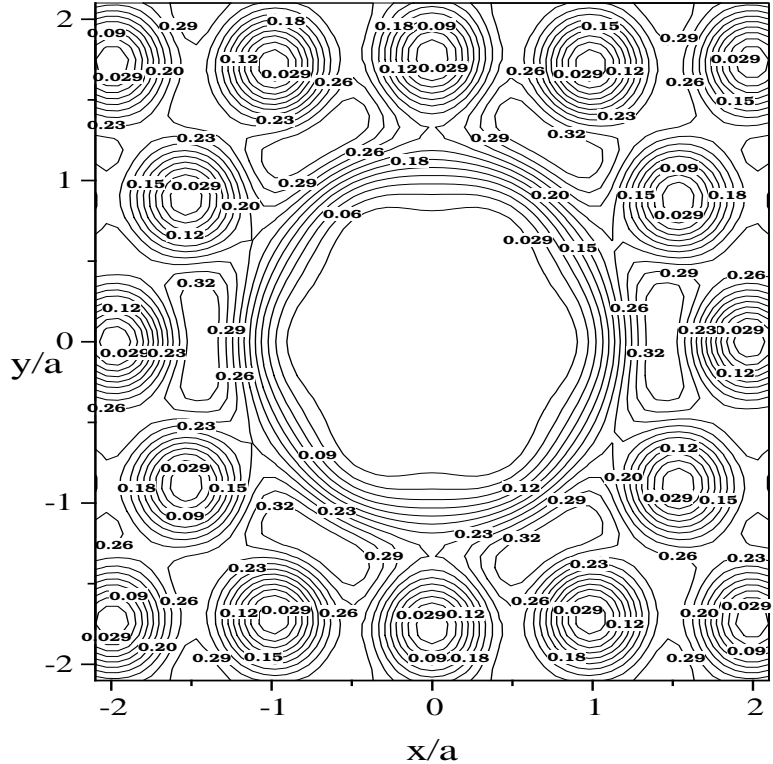}
  \end{center}
  \caption{The dimensionless square modulus of the order parameter
  $\Delta$ in the hexagonal case for parameters $\tilde{\alpha}=0.5,
  \tilde{\gamma}=0.0, \varphi/h =0.5, h=0.93$. 
  Seven vortices collapse on the defect.}
  \label{c6_hol}
\end{figure}
With raising of the applied magnetic field the effective coupling
constants $\alpha$ and $\gamma$ increase drastically
(see Eq.(\ref{scale})), while the parameter $\varphi$ (\ref{radius}) does not 
undergo any visible change. This leads to increasing values of the   
higher coefficients $D(k)$ in the expansion (\ref{C6}) of the order parameter 
and as a result,
to spreading of the deformation far from the defect.The further the growth of the 
magnetic field is, the larger are the  effective 
coupling constants.
This implies that the
last term in the Eq.(\ref{Ovch}) 
 for $k=1$ becomes much larger than all preceding terms. In this case 
the solution is $D_6(1)=-\pi^{-1}M_6(1),$ i.e. the first expansion 
coefficient practically reaches its limiting value. This value  
completely compensates the contribution of the unperturbed Abrikosov 
lattice to the $k=1$ expansion coefficient in Eq.(\ref{C6}). 
In this region of fields the expansion (\ref{C6}) 
begins from $k=7$. The order parameter in the nearest vicinity of 
the 
defect becomes
\begin{eqnarray*}
 \Psi\propto r^7e^{7i\vartheta}.
\end{eqnarray*}

This means that the six nearest vortices have (almost) collapsed on the
defect which pins the vortex containing seven flux quanta.
One can see this effect on fig.\ref{c6_hol}. Here the quantity
\begin{equation}
 |\Delta_6|^2 = \frac{\sqrt{3}|\Psi _6|^{2}}{2\pi |C|^{2}},
\label{delta}
\end{equation}
which is proportional 
to the square modulus of the order parameter (normalization constant $C$ 
is defined by Eq.(\ref{C})), is plotted.\\

With the further increasing the 
applied field the next coefficients $D_6(7),$  $D_6(13)$,  and so on will 
reach 
their limiting compensation values $-\pi^{-1}M_6(7),$  $-\pi^{-1}M_6(13),$ ... , 
and    
one could principally get a vortex 
containing thirteen, nineteen an so on flux quanta. However, numerical 
calculations show that for a realistic field range (not extremely close 
to the upper critical field) only the first collapse 
can be realized.\\

\begin{figure}
 \begin{center}
 \epsfxsize=10.7cm
  \epsfysize=10.4cm
  \epsfbox{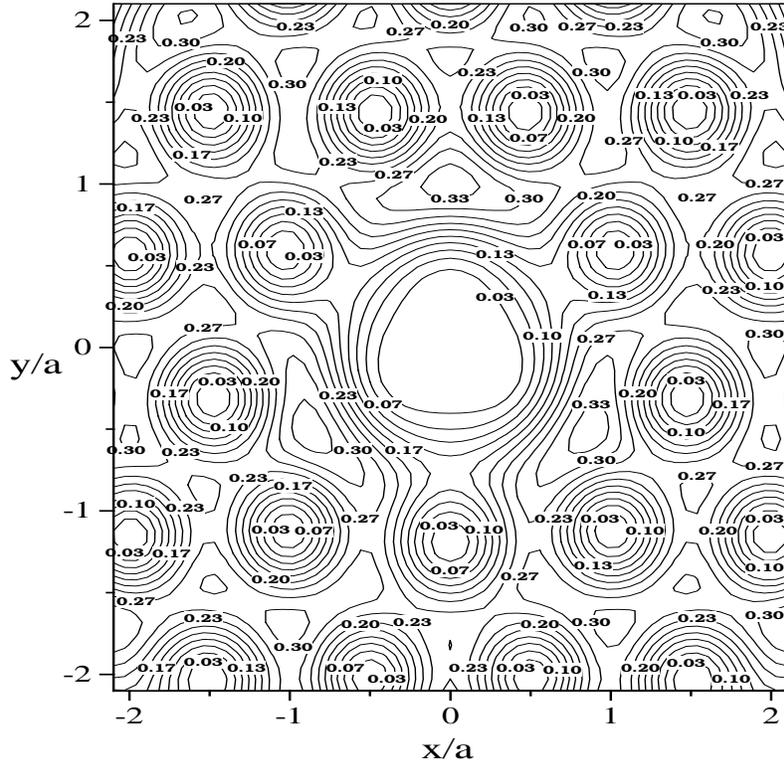}
  \caption{The dimensionless square modulus of the order parameter 
  $\Delta$ in the trigonal case for parameters
  $\tilde{\alpha}=0.5,\tilde{\gamma}=0.0,
  \varphi /h=0.5, h=0.85$. 
  Three vortices collapse on the 
  defect.} 
    \label{c3_hol}
\end{center}
\end{figure}

A similar behavior of the expansion coefficients $\{D(k)\}$ takes place in 
the trigonal case $C_3.$ Here in the case of attraction the coefficient 
$D_3(0)$ is the first one which reaches its compensation value 
$-\pi^{-1}M_3(0),$ that corresponds to the three vortices collapse 
on the defect. Such a configuration is displayed on fig.\ref{c3_hol} where 
the quantity $|\Delta_3|^2$ defined by the r.h.s of 
Eq.(\ref{delta}), with $\Psi_6$ replaced by $\Psi_3$, is plotted 
for the same values of parameters as in the hexagonal case and for the applied 
field $h\approx 0.85$. With increasing of the magnetic field one
expects the appearance of six-, and so on multy-quanta vortices. As in 
the previous case, numerical analysis shows that only the first 
collapse happens in a realistic range of field.\\ 

\begin{figure}
  \begin{center}
  \epsfxsize=11.36cm
  \epsfysize=11.1cm
  \epsfbox{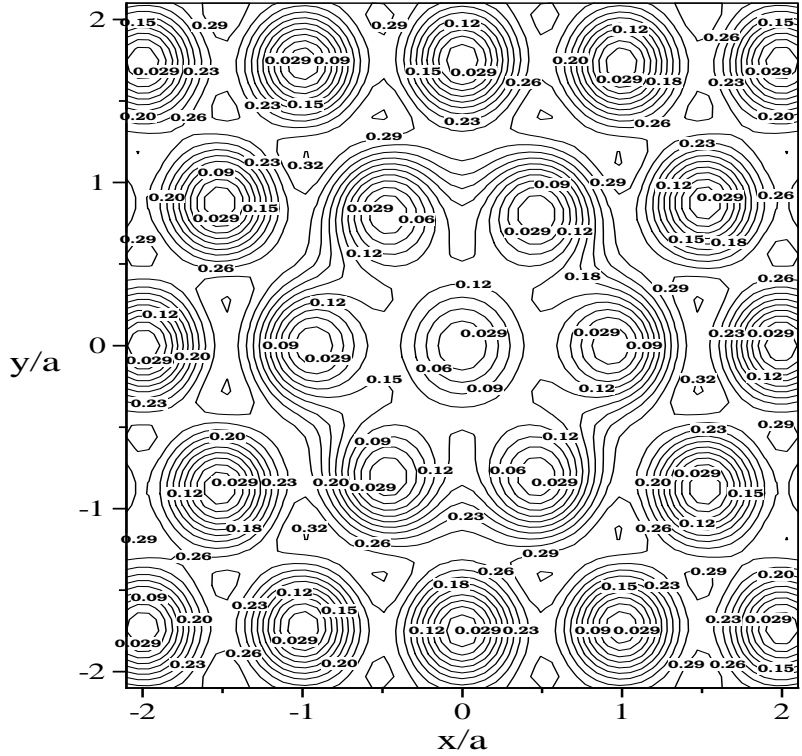}
  \end{center}
  \caption{The dimensionless square modulus of the order parameter 
  $\Delta$ in the hexagonal  case for parameters
  $\tilde{\alpha}=0.5,\tilde{\gamma}=0.0, \varphi /h=0.5, h=0.85$.
  Attractive defect causes a comparatively weak vortex lattice
  deformation.} 
  \label{c6_85}
\end{figure}

Note that for the same set of parameters the first collapse 
within the trigonal symmetry occurs at a weaker field ($h\approx 0.85$) than 
in the hexagonal symmetry ($h\approx 0.93$). The reason is that in the $C_6$ 
system seven vortices must overcome their mutual repulsion in order to fall 
on the defect, while in the $C_3$ system only three vortices collapse. For 
the field $h\approx 0.85$, at which, in the symmetry $C_3,$ three vortices 
are already collapsed on the defect (fig.\ref{c3_hol}), in the $C_6$ symmetry, 
the lattice is distorted but still without any vortex collapse 
(fig.\ref{c6_85}).\\

\begin{figure}
  \begin{center}
  \epsfxsize=13.15cm
  \epsfysize=10.88cm
  \epsfbox{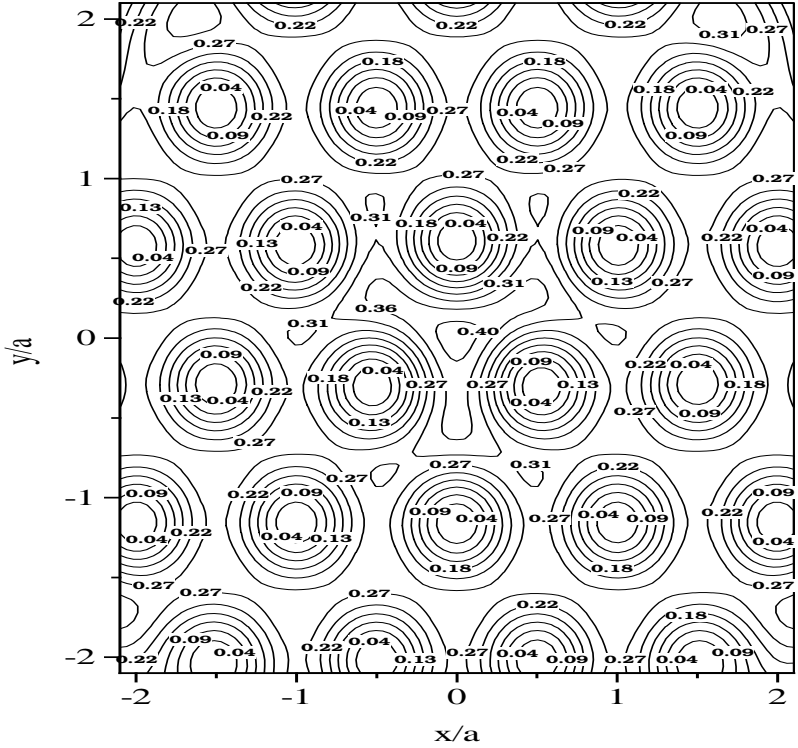}
  \end{center}
  \caption{The dimensionless square modulus of the order parameter 
  $\Delta$ in the 
  hexagonal repulsive case for parameters 
  $\tilde{\alpha}=-0.5,\tilde{\gamma}=0.0,
  \varphi =0.5$.}
  \label{a_}
\end{figure}

The vortex lattice deformation near a repulsive defect is 
presented in fig.{\ref{a_}}.  
Here the three nearest vortices are slightly shifted from the defect
and a visible deformation occurs only in the nearest vicinity of the defect.\\

Up to now we analyzed the solutions of Eqs.(\ref{Ovch}) within two 
symmetries $C_6$ and $C_3$ separately. Now we can choose the most 
preferable one from them and describe the typical vortex lattice behavior 
in some 
interval of the magnetic fields close to the upper critical field. 
We start from the same case of attractive defects $\tilde{\alpha}>0$ 
($\tilde{\gamma}$=0) of a small concentration.
If an applied field is not too close to $H_{c2}$, then a deformation of the 
lattice near a single deffect is small and the preferable local symmetry near 
each defect is $C_6.$  The defects are occupied by vortices
and the rest of the lattice is slightly deformed. With increasing of the 
magnetic field the deformation near defects becomes stronger (as shown in 
fig.\ref{c6_85})
and at some critical field $h_1$ the $C_3$ solution of Eqs.(\ref{Ovch})
corresponding to collapse of three vortices on the defect becomes 
preferable (see fig.\ref{c3_hol}). As a result, a local structural transition  
$C_6$ $\rightarrow$  $C_3$ occurs. With further increasing of the field, one 
deals with $C_3$ symmetry, three vortices occupying the defect and the 
deformation of the nearest (with respect to the defect) part of the vortex 
lattice is observed. But at some critical field $h_2$ the $C_6$ solution of 
 Eqs.(\ref{Ovch}) corresponding to collapse of seven vortices on the defect 
(see fig.\ref{c6_hol}) becomes preferable and a local structural transition  
$C_3$ $\rightarrow$ $C_6$ occurs and so on. Thus, one has a sequence of 
reentering first order phase transitions 
$C_6\rightarrow C_3\rightarrow C_6\rightarrow ...$ .\\

\begin{figure}
  \begin{center}
  \epsfxsize=10.7cm
  \epsfysize=9.46cm
  \epsfbox{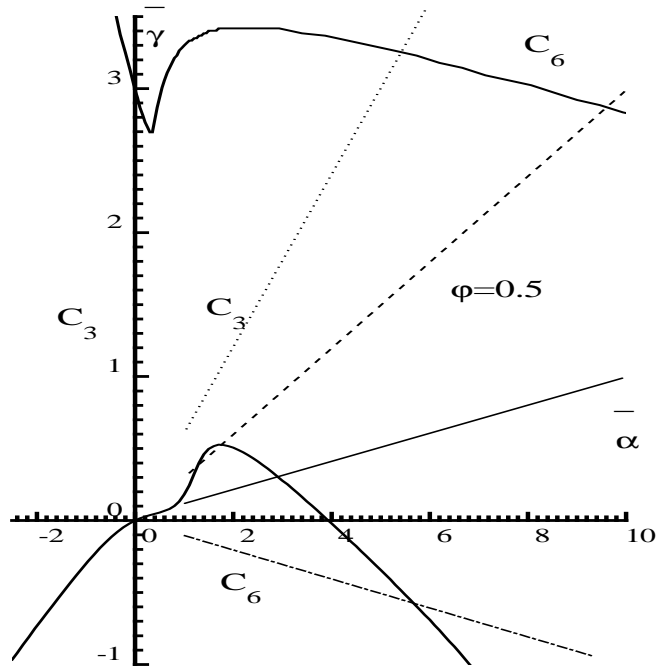}
  \end{center}
  \caption{Phase diagram of superconductor on ($\alpha,\gamma $) plane.
  All rays: $\tilde{\alpha}=0.1; h_0=0.9.$
  Solid ray $\tilde{\gamma}=0.01$.
  Dashed  ray $\tilde{\gamma}=0.03$. 
  Dotted ray $\tilde{\gamma}=0.06$.
  Dashed-dotted ray $\tilde{\gamma}=-0.01.$}
  \label{p0_5p}
\end{figure}

A similar analysis can be done in the general case where 
$\tilde{\alpha}\neq 0$ and $\tilde{\gamma}\neq 0$. 
 The numerical results obtained for various sets of parameters and magnetic 
 field enables us to construct a phase diagram in 
the $(\alpha,\gamma)$ plane 
for a fixed {\it scaled\/} size $\varphi$ (\ref{radius}) of a defect. 
Part of such diagram is given in fig.\ref{p0_5p}. Here the two solid
curves separate the regions where the local symmetry is hexagonal $(C_6)$ or 
trigonal $(C_3)$ . Near the upper critical field $\varphi\approx L^2/\xi^2$
and the diagram becomes universal. For each fixed defect parameters 
and for each value of the magnetic field the diagram enable us to determine the 
preferable local symmetry of the system.\\ 

To explain how to extract this information from the phase diagram consider a 
sample with some fixed parameters $\tilde{\alpha}$, $\tilde{\gamma}$, and 
$L$, and start from an initial applied field 
$h_0\equiv H_0/H_{c2}$. This corresponds to a starting point $(\alpha_0
,\gamma_0)$ in the diagram  of fig.\ref{p0_5p}, where $\alpha_0$ and 
$\gamma_0$ 
are determined by Eqs.(\ref{scale}) with $h=h_0$. Further evolution of the 
parameters $\alpha$ and $\gamma$ with growth of the 
magnetic field is described by equation
$$\gamma=\frac{\tilde{\gamma}}{\tilde{\alpha}}
 \left(\alpha-\alpha _0\right)+\gamma _0$$
and corresponds to some ray on the phase diagram, starting at
the initial point $(\alpha_0,\gamma_0)$ and directed out of the origin. 
Four such rays are displayed in fig.\ref{p0_5p}.  For all rays 
the starting field is $h_0=0.9$ and $\tilde{\alpha}=0.1$. The increasing of 
the magnetic field leads to the change in the effective coupling constants 
(\ref{scale}) i.e. to the motion of a starting point along the 
ray. This movement in its turn results in a sequence of reentering transitions 
from one local symmetry to another.\\ 

The solid ray corresponds to $\tilde{\gamma}=0.01$. It is seen from 
fig.\ref{p0_5p} that at the initial field $h_0=0.9$ the local 
deformation has a $C_6$ symmetry. This is consistent with the 
analytical prediction of Section III: the deformation around a defect is 
small and the inequality (\ref{ineq}) is valid.
As the field increases, the ray crosses the lower solid curve and
the sample undergoes a first order phase transition to the trigonal local  
symmetry $C_3$. At this symmetry we have three collapsed vortices 
at each defect. Transition to the symmetry $C_6$ back is also possible, but 
it is not seen on the diagram because it occurs  in the region $\alpha\gg 
1$ i.e. at a field extremely close to the upper critical field $H_{c2}$.\\

The dashed ray corresponds to the value $\tilde{\gamma}=0.03$ and 
represents probably the most interesting case. 
Here even in the comparatively low field $h\approx 0.775<h_0$ 
(the corresponding point of the ray is not displayed on 
fig.\ref{p0_5p}) the $C_6$-$C_3$ symmetry transition occurs. In both the two 
lattice configurations below and above the transition the lattice deformation 
 is small and can be described within the approach of Section III. 
The inequality (\ref{ineq}) is violated below the transition field and is 
 valid above it. 
The dashed ray on the diagram starts from the field $h_0=0.9$ and for the 
first time crosses  
the lower solid curve at a field $h\approx 0.906$, at 
which the lattice undergoes the next $C_3$ $\rightarrow$ $C_6$ transition. 
No vortex collapse still happens at this field because
the value of $D_6(1)$ is still far from its compensating value. However two 
next transitions take place because of vortex collapse. The second transition 
to the symmetry $C_3$ at a field $h\approx 0.94$ 
happens when the coefficient 
$D_3(0)$ in the symmetry $C_3$ almost reaches its compensating value 
$D_3(0)=-\pi^{-1}M_3(0)$ 
and therefore this transition corresponds to the collapse 
of the three vortices at the defect. 
Similarly the third transition 
to the 
symmetry $C_6$ at a field 
$h\approx 0.99$  corresponds to the collapse of the seven vortices at the 
defect.
\\ 

The dotted ray in fig.\ref{p0_5p} corresponds to $\tilde
{\gamma}=0.06$ that provides only a $C_3$ symmetry in the comparatively low 
field region. In the high field region we obtain $C_3$ $\rightarrow$ $C_6$ 
transition at the field $h\approx 0.98$.\\ 

Note that the figures \ref{c3_hol} and \ref{c6_85} already reffered to above, 
present 
the contour plots of the order parameter near defect in the vicinity of the 
$C_6$ $\rightarrow$ $C_3$ 
transition due to collapse of the three nearest vortices on the defect. These 
plots correspond to the point $(\approx 4, 0)$ on the ray coinciding with 
the positive $\alpha-$semiaxis on the phase diagram. At this point the order parameter 
exhibits a small deformation in the symmetry $C_6$ as it is displayed in 
fig.\ref{c6_85}, while in the symmetry $C_3$ it is strongly deformed due to the collapse 
(see fig. \ref{c3_hol}).\\

In the region where $\alpha>0$ and $\gamma<0$ a local symmetry 
transition $C_6\rightarrow C_3$ due to vortex collapse is described by  the 
last fourth ray on 
the diagram. This ray corresponds to the parameters 
$\tilde{\alpha}=0.1$, $\tilde{\gamma}=-0.01$.\\

\section{Small concentration of the defects}
During the two previous sections we dealt with a single defect problem. To 
be sure that our results (\ref{F}) do describe a macroscopic system with a finite 
concentration of the defects we have to be sure that the next (second order) 
concentration correction to the thermodynamic potential is small. To estimate 
this correction one has to solve exactly the two defects problem which is 
much more complicated. Therefore we choose another way.\\  

Consider for simplicity an attractive case and magnetic field which is
not too close to $H_{c2}.$ Put the undistorted vortex lattice on
the plane where (point) defects are distributed and shift one of the
vortices nearest to each inhomogeneity to the position of that
inhomogeneity. There are many similar ways to arrange the vortex lattice,
but one has to choose such a way which leads to alternation of the regions
where the lattice is compressed with ones where its rarefied. Finally let
us distort the regions of the lattice close to inhomogeneities according
to the results obtained within single defect approximation. This latter
distortion is already taken into account {\it exactly}. So one has only to
estimate the additional contribution to the thermodynamic potential from 
the intermediate regions (between inhomogeneities) whose deformation is well 
described by elastic theory.\\
 
The number of extra vortices per region is of order of unity. Therefore the 
deformation tensorup to a numerical
factor of the order of unity equals to dimensionless concentration of defects
$c$. The correction to the thermodynamic potential will be of the order of
$Cc^2$, where $C$ is the elastic modulus. But the elastic part of the
deformation has an alternating behavior with a characteristic wavelength
of the order of the average distance between inhomogeneities. As it was
shown by E. Brandt \cite{Brandt1}, all the elastic moduli are proportional
to $(1-H/H_{c2})^2$ if this distance is much less than the penetration
length divided by $(1-H/H_{c2})^{1/2}$. The latter inequality can be
rewritten as $36\kappa c\gg 1-H/H_{c2}$, where $\kappa\gg 1$ is the
Ginzburg-Landau parameter. In the region of parameters which we are mostly
interested in $c=0.03$, $1-H/H_{c2}=0.06$ and the inequality $\kappa\gg 1$
is evidently valid. This means that corresponding contribution to the
thermodynamic potential is of the order of $(1-H/H_{c2})^2c^2$. This is exactly 
the second order concentration correction which in the case $c\ll 1$ is
smaller than the contribution accounted for within the linear
concentration expansion.\\

Thus in the case of small concentration one can use the results obtained in 
the two previous sections and describe the thermodynamics of the system 
near $H_{c2}$. Define a dimensionless magnetization
$$m\equiv -4\pi(2\kappa^2-1)\beta_A \frac{M}{H_{c2}}$$
 and dimensionless magnetic succeptibility
$$\chi=\frac{\partial m}{\partial h}.$$

\begin{figure}
  \begin{center}
  \epsfxsize=10.7cm
  \epsfysize=9.14cm
  \epsfbox{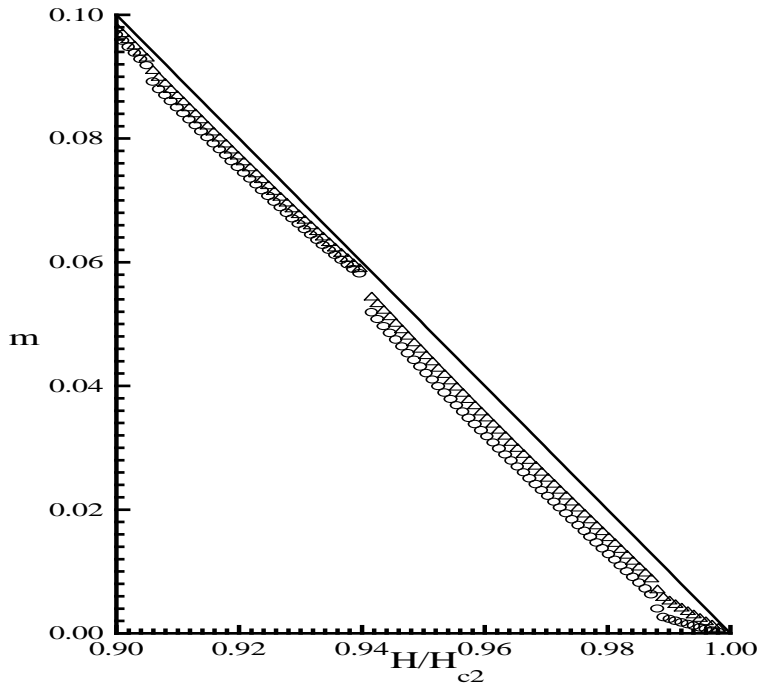}
  \caption{The magnetization curve of superconductor for the parameters
  $\tilde{alpha}=0.1, \tilde{\gamma}=0.03$ and for the 
  concentration values $c=0.03$ (triangles) and $c=0.05$ (circles).}
  \label{p0_5m}
\end{center}
\end{figure}
\begin{figure}
  \begin{center}
  \epsfxsize=10cm
  \epsfysize=8.88cm
  \epsfbox{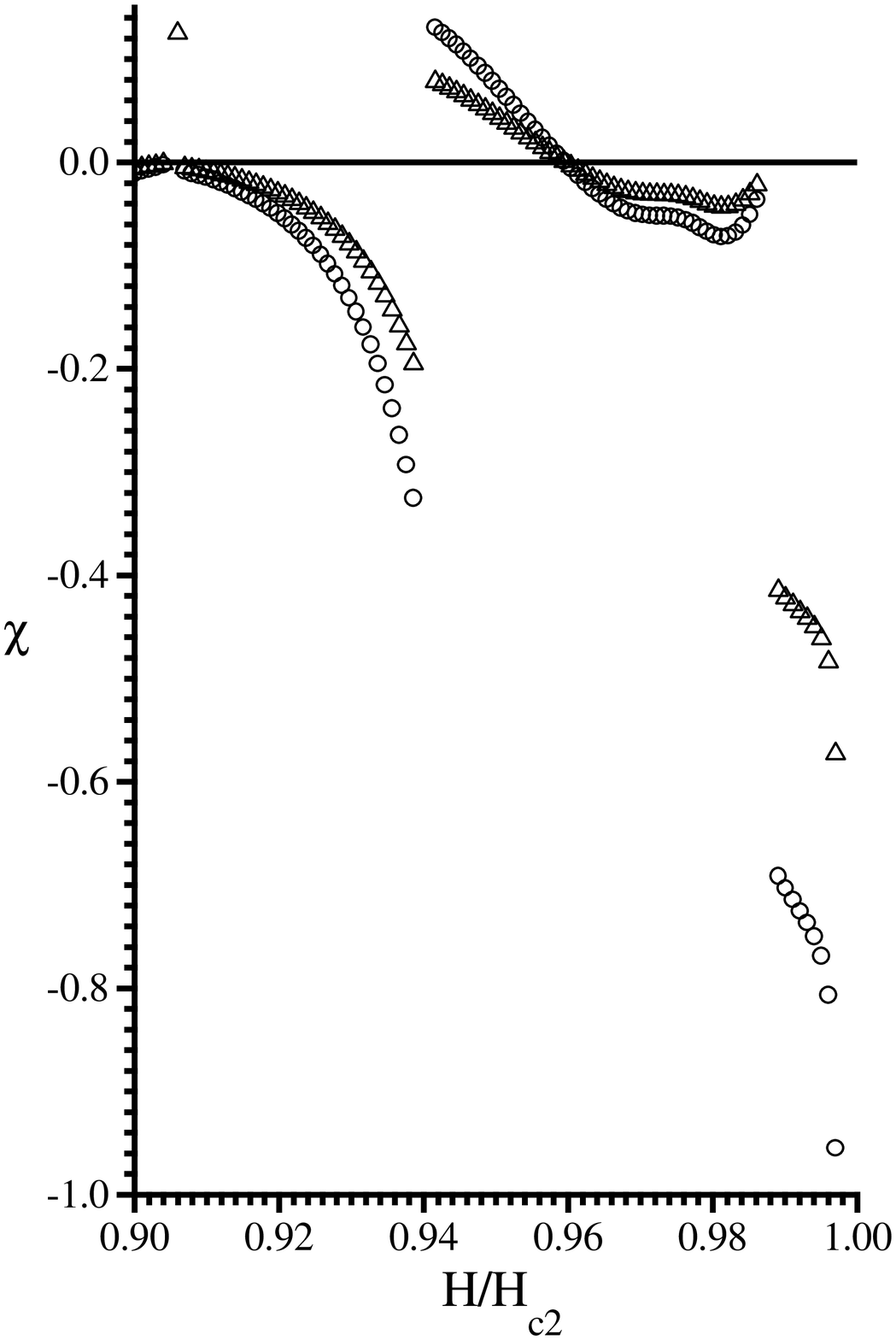}
  \caption{The magnetic succeptibility of superconductor for the
  parameters $\tilde{alpha}=0.1,\tilde{\gamma}=0.03$ and for
  the concentration values $c=0.03$ (triangles) and $c=0.05$ (circles).}
  \label{p0_5s}
\end{center}
\end{figure}

 All the local symmetry transitions described above manifest 
themselves as jumps on the magnetization curve (fig.\ref{p0_5m}) and as 
discontinuities on the magnetic succeptibility 
curve (fig.\ref{p0_5s}). 
The most pronounced jumps occur at the two transitions accompanied by  
vortex collapse, namely at the fields $h\approx 0.94$ and $h\approx 
0.99$.\\

\section{Summary}
We studied the equilibrium properties of conventional type II superconductor 
with small concentration of randomly placed identical columnar defects. 
In the vicinity of the upper critical field the vortex lattice undergoes a 
strong deformation with two possible local symmetries -- hexagonal one $C_6$
 and trigonal one $C_3$. The character of the 
deformation is determined by the vortex-defect interaction. The vortices can 
collapse onto attractive defects and the formation of multy-quanta vortices 
becomes possible. Formation of the multiquantum vortices was predicted 
earlier \cite{Shapiro}, but in "twice" opposite limiting case. We deal 
with a short-range defect and gain an energy because of softening of the 
Abrikosov lattice near $H_{c2},$ while in  \cite{Shapiro} a  very strong 
defect with a radius comparable with the penetration length was 
considered.\\   

Increasing the external field gives rise to the reentering 
transitions between the two possible types of symmetry. These transitions can 
be described by a universal phase diagram. They manifest themselves as jumps 
of the  magnetization and peculiarities of the magnetic susceptibility.

One of the way to observe these equilibrium states near $H_{c2}$ is to cool 
a sample subject to a magnetic field in the normal state, below the 
critical temperature. Another possibility is to observe not the 
equilibrium state as a whole, but visualize the local deformation of the vortex 
lattice near defects.

\section{Acknowledgements}
This study was partially supported by the Israel Science Foundation. 
We are grateful to H. Brandt, E. Chudnovsky, B. Horovitz, V. Mineev, R.
Mints, Z. Ovadyahu, B.Ya. Shapiro, and E. Zeldov for very helpful discussions 
of our results.\\

\appendix
\section{Abrikosov Lattice Expansion}
\label{exp_teo}
The Abrikosov order parameter  \cite{Abrik} 
$$\Psi^A(x,y)=C\exp\left[-\frac{2\pi}{\sqrt{3}}\frac{y^2}{a^2}\right]
 \theta_3\left(\left.\frac{x}{a}-i\frac{y}{a}-\frac{1}{4}\right|
 e^{i\frac{\pi}{3}}\right),$$
which was obtained in the Landau gauge ${\bf A}=\hat{x}H(-y,0)$
describes a triangular vortex lattice with sites
\begin{eqnarray*}
 \left\{
 \begin{array}{c}
 x=a(2m+n+2)/2,\\
 y=a\sqrt{3}(2n+1)/4,
 \end{array}
 \right.
\end{eqnarray*}
 ($m$ and $n$ are integers). Here 
\begin{equation}
	a(T)=2\xi (T)\sqrt{\frac{\pi}{h(T)\sqrt{3}}}
	\label{latconst}
\end{equation}
is the triangle side (Abrikosov lattice constant), and $h(T)\equiv H/H_{c2}(T).$
The  
Abrikosov normalization constant $C$ is
related to the 
thermodynamic potential density (\ref{FE}) of the clean superconductor by
\begin{equation}
	|C|^{2}=\frac{3^{1/4} 2|f^{A}|}{|\alpha _0|(1-h)},
	\label{C}
\end{equation} 
and $\theta_3(u|\tau)$ is the Euler $\theta$ -function \cite{GRy}.\\

The order parameter
$$\Psi_{x_0,y_0}(x,y )\equiv
 \exp\left[
  -i\frac{2\pi}{\sqrt{3}a^2}x y+
  i\frac{4\pi y_0}{\sqrt{3}a}x
 \right]
 \Psi^A (x -ax_0,y -ay_0),$$
describes the shifted Abrikosov lattice in the symmetric gauge. This order 
parameter can be expanded as 
$$\Psi_{x_0,y_0}(x,y)=\pi^{-1}\sum_{k=0}^{\infty}M_{x_0,y_0}(k)
  L_k(x,y)$$
with respect to Landau functions with the orbital moment $k\geq 0$ 
\begin{equation}
 L_k(x,y)=\frac{C}{3^{1/4}}\sqrt{\frac{2\pi}{k!}}
 \left[\frac{\sqrt{2\pi}}{3^{1/4}}\frac{r}{a}\right]^k
 \exp\left[ -ik\vartheta -\frac{\pi r^2}{\sqrt{3}a^2}\right],
 \label{Land}
\end{equation}
of the lowest Landau level of 
a particle with electron mass and charge $-2e$ in the 
magnetic field  $H.$ 
The expansion coefficients are
\begin{eqnarray*}
 M_{x_0,y_0}(k)=\sqrt{\frac{\sqrt{3}\pi}{2}}
 \sum_{n=-\infty}^{\infty}\sum_{m=0}^{k-2m\geq 0}
 \frac{i^{k-2m}\left[\sqrt{2\pi\sqrt{3}}\right]^{k-2m}}{2^mm!(k-2m)!}
 \left(n+\frac{2y_0}{\sqrt{3}}\right)^{k-2m}\times\nonumber\\
 \exp\left\{
 i\frac{\pi}{2}\left[
 n^2-n(1+4x_0)
 \right] -\frac{\pi\sqrt{3}}{2}\left(
 n+\frac{2y_0}{\sqrt{3}}
 \right)^2
 \right\}.
\end{eqnarray*}

The case $x_0=-1/2$, $y_0=-\sqrt{3}/4$ corresponds to the $C_6$ symmetry 
when one vortex is placed at the origin. Taking into account the selection 
rule (\ref{selectionM}) we write down the expansion coefficients as
\begin{eqnarray*}
 M_{-1/2,-\sqrt{3}/4}(k)=i\delta_{k,6K+1}M_6(6K+1),
\end{eqnarray*}
where 
\begin{eqnarray}
 M_6(k)=\sqrt{\frac{\sqrt{3}\pi}{2}}
 \sum_{n=-\infty}^{\infty}\sum_{m=0}^{k-2m\geq 0}
 \frac{i^{k-2m-1}\left[\sqrt{2\pi\sqrt{3}}\right]^{k-2m}}{2^mm!(k-2m)!}
 \left(n-\frac{1}{2}\right)^{k-2m}\times\nonumber\\
 \exp\left\{
 i\frac{\pi}{2}\left[
 n^2+n
 \right] -\frac{\pi\sqrt{3}}{2}\left(
 n-\frac{1}{2}
 \right)^2
 \right\}.
\label{M6}
\end{eqnarray}
The expansion coefficients $M_6(k)$ are real. The values of the first few of 
them are contained in the table I.\\ 
\begin{table}
\caption{Values of an Expansion Coefficients $M_6(k)$}
\begin{tabular}{ll}
$k$ & $M_6(k)$\\
\tableline
$1$&$-2.792$\\
$7$&$4.057$\\
$13$&$2.260$\\
$19$&$-4.852$\\
$25$&$-1.494$\\
$31$&$-3.538$\\
$37$&$4.817$\\
$43$&$2.605$\\
$49$&$0.479$\\
$55$&$5.180$\\
\end{tabular}
\end{table}
The second case $x_0=-1/2$ and $y_0=\sqrt{3}/12$ corresponds to the $C_3$ 
symmetry, when the origin of the coordinate system is placed in the 
center of an elementary triangle. Selection rules allow us to write 
down the expansion coefficients as
\begin{eqnarray*}
 M_{-1/2,\sqrt{3}/12}(k)=i^{-3K}\delta_{k,3K}M_3(3K).
\end{eqnarray*}
\newpage
The explicit expression for $M_3(k)$ is given by the formula
\begin{eqnarray}
 M_3(k)=\sqrt{\frac{\sqrt{3}\pi}{2}}
 \sum_{n=-\infty}^{\infty}\sum_{m=0}^{k-2m\geq 0}
 \frac{i^{2k-2m}\left[\sqrt{2\pi\sqrt{3}}\right]^{k-2m}}{2^mm!(k-2m)!}
 \left(n+\frac{1}{6}\right)^{k-2m}\times\nonumber\\
 \exp\left\{
 i\frac{\pi}{2}\left[
 n^2+n
 \right] -\frac{\pi\sqrt{3}}{2}\left(
 n+\frac{1}{6}
 \right)^2
 \right\}.
\label{M3}
\end{eqnarray}
All the coefficients $M_3(k)$ also are real. The values of the first few coefficients are given 
by Table II.
\begin{table}
\caption{Values of an Expansion Coefficients $M_3(k)$}
\begin{tabular}{ll}
$k$ & $M_3(k)$\\
\tableline
$0$&$1.738$\\
$3$&$-2.942$\\
$6$&$-2.227$\\
$9$&$1.646$\\
$12$&$-3.568$\\
$15$&$-2.310$\\
$18$&$-0.756$\\
$21$&$3.185$\\
$24$&$-1.563$\\
$27$&$3.110$\\
$30$&$3.606$\\
\end{tabular}
\end{table}

\end{document}